\documentclass{aa}  

\usepackage{graphicx}
\usepackage{txfonts}
\usepackage[colorlinks,citecolor=blue]{hyperref}
\usepackage[rgb]{xcolor}
\usepackage[normalem]{ulem}
\usepackage{enumitem}
\usepackage{float}

\usepackage{orcidlink}

\def\sgr{Sgr~A$^{\star}$}
\def\rh{r_{\rm H}}

\begin{document} 

\title{Emergence of cyclic flux eruptions in kinetic simulations of magnetized spherical accretion onto a Schwarzschild black hole}

\titlerunning{Cyclic flux eruptions in kinetic simulations of magnetized spherical accretion}

   \author{Enzo Figueiredo\inst{1}\orcidlink{0009-0006-7103-1965}
          \and
          John Mehlhaff\inst{1,2}\orcidlink{0000-0002-7414-0175}
          \and
          Adrien Soudais\inst{1,3}\orcidlink{0000-0002-8599-8847}
          \and
          Beno\^it Cerutti\inst{1}\orcidlink{0000-0001-6295-596X}
          }

\institute{Univ. Grenoble Alpes, CNRS, IPAG, 38000 Grenoble, France \and Physics Department and McDonnell Center for the Space Sciences, Washington University in St.\ Louis; MO, 63130, USA \and Department of Physics and Astronomy, Dartmouth College, Hanover, NH 03755, USA
          \\
           \email{enzo.figueiredo@univ-grenoble-alpes.fr}
           }

\date{Accepted June 13, 2026}

  \abstract
   {The dynamics of black hole magnetospheres critically depend on the black hole spin and on the structure of the accretion flow. In the limit of a Schwarzschild black hole immersed in a zero-net angular momentum flow, accretion is spherical. However, in the presence of a large-scale vertical magnetic field, the classical Bondi accretion model is significantly altered. The frozen-in field is stretched radially as the plasma is pulled inward by gravity. This continues until the restoring force from the magnetic tension suddenly expels the material and resets the field, allowing a new cycle to begin.}
   {Although this scenario has been well depicted in previous studies, it remains incomplete as the issues of dissipation and particle acceleration are not yet fully resolved. In this work, we aim to revisit these issues with a first-principles kinetic plasma model.}
   {We perform two-dimensional global general relativistic particle-in-cell simulations of magnetized spherical accretion onto a Schwarzschild black hole, for both pair and electron-ion plasmas. The simulations are evolved over long timescales to capture multiple flux eruption events and establish a quasi-steady state.}
   {For each accretion cycle, we find that the system goes through three main stages: (i) an ideal advection phase where magnetic flux through the horizon increases quasi-linearly with time; (ii) a reconnection-regulated phase where the net increase of the flux is slowed down by intermittent reconnection events near the horizon; and (iii) a flaring phase when a major, large-scale reconnection event expels the flux, leading to efficient particle acceleration.}
   {The emergence of large-amplitude quasi-periodic flux eruptions and concomitant particle acceleration is reminiscent of Sgr~A$^{\star}$ flaring activity. This phenomenon could also be applicable to quiescent black holes, especially isolated black holes accreting the interstellar medium.}


   \maketitle
%
%-------------------------------------------------------------------
\section{Introduction}
Accreting black holes are frequently associated with the launching of powerful relativistic jets, originating from the innermost regions of the accretion flow and possibly powered by the black hole ergosphere (see \citealt{2019ARA&A..57..467B} for a review and references therein). In high-luminosity systems, the accretion rate is large enough for the flow to cool efficiently, resulting in a geometrically thin, optically thick Keplerian disk \citep{1973A&A....24..337S}. In contrast, at low accretion rates, radiative cooling becomes inefficient, and the accretion flow becomes hot, geometrically thick, and more tenuous \citep{1994ApJ...428L..13N}. In fact, the plasma density is sufficiently low that the flow becomes effectively collisionless, with Coulomb collision times exceeding the accretion timescale (e.g., \citealt{2002ApJ...577..524Q}).

This regime is particularly relevant for the nearby supermassive black holes M87$^{\star}$ and \sgr, where recent horizon-scale observations provide some of the most stringent empirical constraints on accretion physics. A key conclusion inferred from polarimetric measurements of these two sources is the presence of an organized vertical magnetic field on large scales \citep{2015Sci...350.1242J, GRAVITY2018, 2020A&A...635A.143G, 2021ApJ...910L..13E, eht_etal_2024, 2022A&A...665L...6W}. Such a dynamically important magnetic field can dramatically alter the nature of accretion by exerting pressure and stress on the flow. Although \sgr{} does not exhibit a detectable jet, it nevertheless displays puzzling activity characterized by frequent nonthermal flares originating from the immediate vicinity of the black hole horizon \citep{baganoff_etal_2001, genzel_etal_2003}, which indicates ongoing particle acceleration \citep{2001A&A...379L..13M, 2010ApJ...725..450D}.

In \sgr, the accretion flow most likely results from the collective contributions of massive star winds orbiting the black hole \citep{2006ApJ...643.1011P}. Dedicated studies show that the plasma penetrates the black hole vicinity with a low net angular momentum \citep{2005MNRAS.360L..55C, 2018MNRAS.478.3544R}, which in turn suggests that accretion may be close to spherical. In the presence of a large-scale vertical magnetic field, the classical Bondi accretion model is significantly altered \citep{bisnovatyi-kogan_ruzmaikin, 2002ApJ...566..137I, 2021MNRAS.504.6076R}. The frozen-in magnetic field lines are dragged inward with the plasma as it is gravitationally pulled towards the black hole event horizon. As accretion proceeds, field lines become increasingly stretched in the radial direction, leading to a partial storage of the gravitational potential energy in the field.

This process continues until the field becomes dynamically important near the hole. At this stage, the restoring force from the magnetic tension pushes the accretion flow away from the horizon. This phenomenon is accompanied by a large-scale reconnection event that facilitates the global reorganization of the field and yields efficient non-thermal particle acceleration \citep{bisnovatyi-kogan_ruzmaikin, 1976Ap&SS..42..401B, 1975A&A....44...59M}. This is reminiscent of flux eruptions reported in general relativistic magnetohydrodynamic (GRMHD) simulations of magnetically arrested disks (e.g., \citealt{2011MNRAS.418L..79T, 2020MNRAS.497.4999D, 2021MNRAS.502.2023P, 2022ApJ...924L..32R}). This scenario provides a promising explanation for the recurrent flares observed in \sgr{} mentioned earlier. Nevertheless, this picture remains incomplete because the questions of dissipation and particle acceleration have not yet been fully settled. Fluid models have been widely used to elaborate the above picture, but they are unable to capture the microphysics of collisionless plasmas and particle acceleration. A kinetic plasma approach must therefore be employed.

Recent global general relativistic particle-in-cell (GRPIC) simulations of magnetized spherical accretion around a maximally spinning black hole have uncovered a clear connection between flux eruptions and particle acceleration \citep{2023PhRvL.130k5201G, 2025PhRvL.135a5201V}. Building on these previous works, we present here a new series of two-dimensional GRPIC simulations of collisionless spherical accretion for a Schwarzschild black hole immersed in an initially vertical laminar field (Sect.~\ref{sec:Methods}). This configuration, by design, prevents the formation of a spin-powered jet \citep{1977MNRAS.179..433B}, in contrast to previous GRPIC studies, and thus focuses on the accretion process. Solutions are integrated over an unprecedentedly long timescale so that a quasi-steady state can be reached and multiple flux eruptions can be observed in the presence of a pair plasma (Sect.~\ref{sec:Pairs}) or an electron-ion plasma (Sect.~\ref{sec:Ions}). We further dissect each stage of the accretion process and propose a simplified model to construct a quantitative physical framework based on the simulation results (Sect.~\ref{sec:Model}). Results are then discussed in the context of \sgr{} flares and isolated stellar-mass black holes accreting from the interstellar medium (Sect.~\ref{sec:Discussion}).

%--------------------------------------------------------------------

\section{Methods} \label{sec:Methods}

\begin{figure}[h]
    \centering
    \includegraphics[width=\columnwidth]{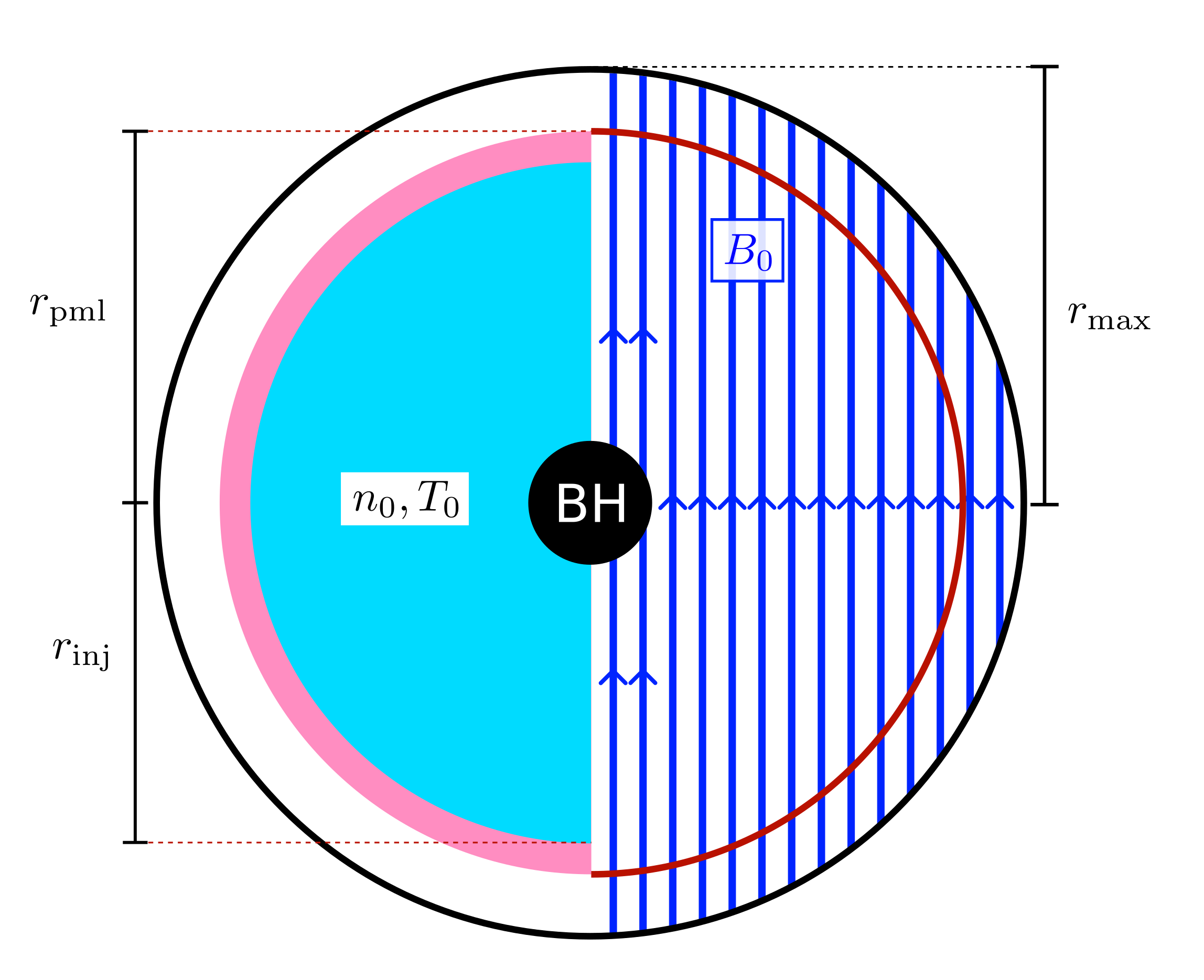}
    \caption{Sketch of the initial GRPIC setup representing zero-angular-momentum accretion onto a central black hole (labeled BH) in the center. On the left, we show in cyan the initial cloud of plasma, at a density $n_0$ and temperature $T_0$, and in pink the injection area of fresh plasma. On the right, the initial uniform magnetic field lines are shown in blue, and the limit of the outer matching layer in red.}
    \label{fig:sketch}
\end{figure}

In this work, we use the GRPIC code \texttt{Zeltron} \citep{Parfrey2019}, where the fields and particle properties are derived in the 3+1 formalism of \citet{Komissarov2004} using spherical Kerr-Schild coordinates. In this formalism, $\textbf{B}$ and $\textbf{D}$ are the magnetic and electric fields as measured by fiducial observers (FIDOs) and $\textbf{H}$ and $\textbf{E}$ are auxiliary, metric-induced magnetic and electric fields.
We model zero-angular-momentum accretion of matter by a Schwarzschild black hole of radius $r_{\rm H} = 2 r_g$, where $r_g = \mathcal{G}M_{\rm BH}/c^2$ is the gravitational radius and $M_{\rm BH}$ is the black hole mass. In this article, we employ Gaussian-cgs units for the electromagnetic fields and use~$r_g$ and~$t_g=r_g/c$ as explicit units of space and time, but we otherwise set~$\mathcal{G}=c=M_{\rm BH}=1$.

We employ a 2D spherical grid of size $N_r \times N_\theta$. The grid is logarithmically spaced in the radial direction and uniform in~$\theta$, and it covers the domain $[r_{\rm min},r_{\rm max}]\times[0.013\pi,0.987\pi]$, where $r_{\rm min}<r_{\rm H}$. We enforce axial symmetry at the $\theta$ boundaries.
Waves and particles are absorbed in a layer between $r_{\rm pml} = 30 r_g$ and~$r_{\rm max} = r_{\rm pml} / 0.9 \simeq 33 r_g$. The magnetic field lines in this layer are matched to the initial field.

The initial electromagnetic configuration follows the \citet{wald1974} solution with zero spin, corresponding to a uniform magnetic field at infinity of intensity $B_0$. We assume that the black hole is initially surrounded by a stationary homogeneous cloud of thermal plasma with total number density $n_0$ and temperature $T_0$ as measured by FIDOs. The plasma is composed of electrons of mass $m_e$ and charge $-e$ and ions of mass $m_i$ and charge $+e$, both with equal number densities~$n_0/2$. To maintain a supply of fresh infalling plasma, we inject electrons and ions in a thin spherical shell between $r_{\rm inj} = r_{\rm pml} - r_g$ and $r_{\rm pml}$, enforcing a local number density floor of $n \geq n_0$. This injected plasma is thermal with the same temperature~$T_0$ as the initial cloud. We enforce roughly $N_{\rm PPC}=10$ particles per cell per species both throughout the domain at the simulation onset and at later times in the injection layer. We checked that the results are not quantitatively affected up to $N_{\rm PPC}=50$. Our initial and boundary conditions are summarized in Fig.~\ref{fig:sketch}.

Accretion can only occur if the ion thermal velocity,~$v_{\rm th} = \sqrt{k_B T_0/m_i}$, where $k_B$ is the Boltzmann constant, is lower than the gravitational escape velocity,~$v_{\rm esc} = \sqrt{2r_g /r}$. This imposes a constraint on the normalized temperature,~$\vartheta_0 \equiv k_B T_0 / m_i$, of~$\vartheta_0 \leq 2 r_g / r_{\rm pml}$. To respect this limit, we set~$\vartheta_0 = 1/30$, which also means that the Bondi radius $R_{\rm Bondi} = 2\mathcal{G}M_{\rm BH}/v^2_{\rm th}$ lies outside of our simulation box. In practice, we set the species-specific normalized temperatures,~$\vartheta_{0,e} = k_{\rm B} T_{0,e} / m_e$ and~$\vartheta_{0,i} = k_B T_{0,i} / m_i$, both equal to~$\vartheta_0 = k_{\rm B} T_0 / m_i$. For~$m_i \neq m_e$, this renders electrons initially colder than ions, but only reduces the overall plasma temperature from~$T_0$ by at most a factor of two.

We target a low initial magnetization $\sigma_0 = B^2_0 / 4 \pi \rho_0 < 1$, where $\rho_0 = (m_i + m_e) n_0/2$ is the plasma mass density. If~$\sigma_0$ were too high in the injection zone, matter would not accrete due to the rigidity of magnetic field lines. We vary~$\sigma_0$ for different simulations, but it is always of order~$0.1$. This also means that the plasma beta-parameter is $\beta_{\rm plasma} =  8 \pi n_0 k_B T /B^2 \gtrsim 1$. 

\begin{figure*}
\centering
   \includegraphics[width=\textwidth]{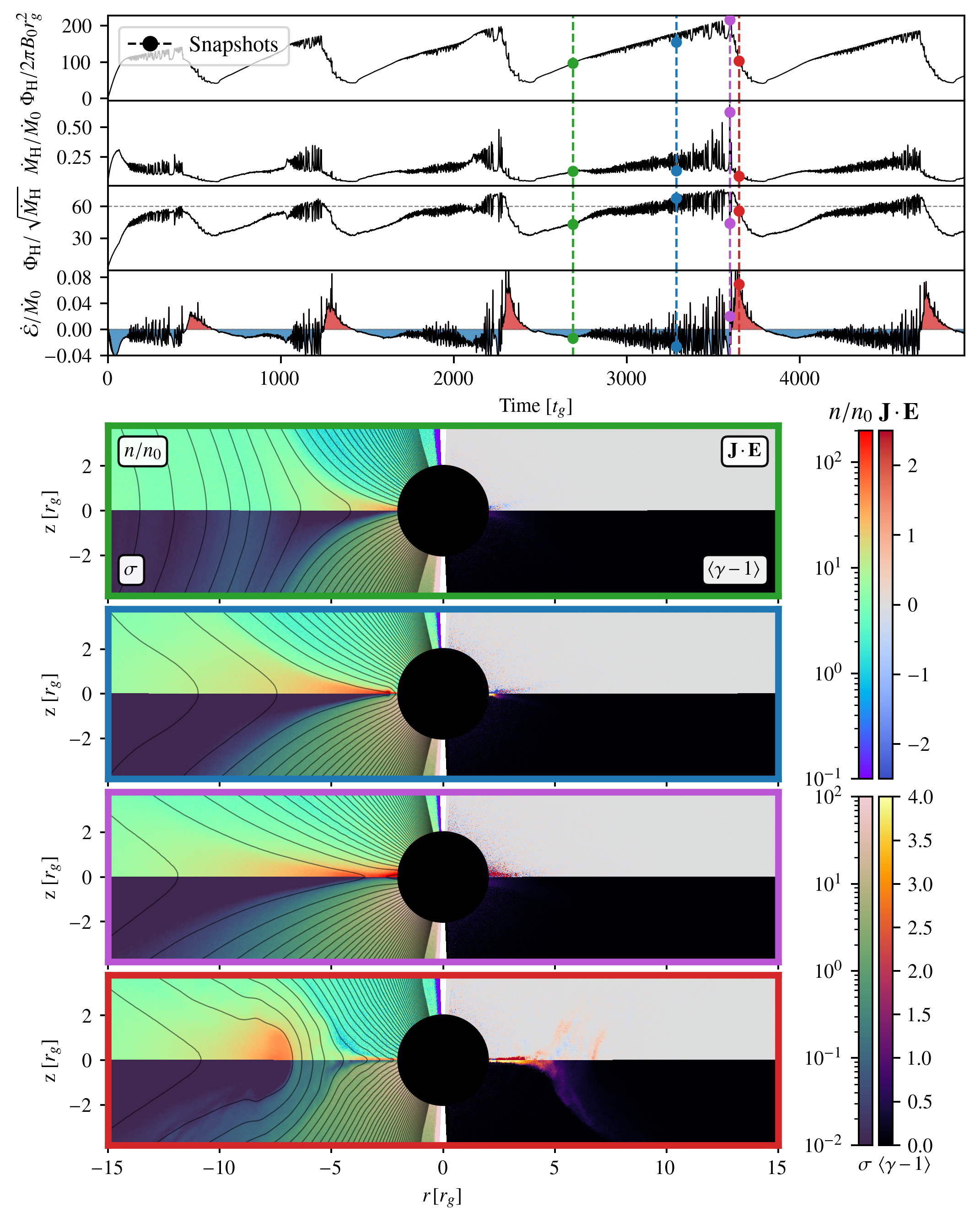}
     \caption{\textbf{Top half}: Time evolution of: the horizon magnetic flux, $\Phi_{\rm H}$; the horizon accretion rate, $\dot{M}$, normalized by $\dot{M}_{0}=4\pi m_i n_0 r_{\rm inj}^2 v_{\rm th}$; the normalized horizon magnetic flux, $\Phi_{\rm H}/\sqrt{\dot{M}}$; and the total dissipation rate, $\dot{\mathcal{E}}$, normalized by $\dot{M}_{0}$. \textbf{Bottom half}: Simulation snapshots at~$t/t_g=2689, 3286, 3596,\,\rm and\, 3648$. Each snapshot shows the number density $n$ (top left), the electromagnetic energy dissipation $\mathbf{J}\cdot\mathbf{E}$ (top right), the magnetization $\sigma$ (bottom left) and the local average particle Lorentz factor $\langle\gamma-1\rangle$ (bottom right). Black lines show magnetic field lines.}
     \label{fig:bigfig}
\end{figure*}

The main parameter we explore is the ion-to-electron mass ratio,~$m_i/m_e$. A realistic mass ratio of~$m_i/m_e = 1836$ imposes a broad separation between ion and electron scales and is therefore too computationally expensive to achieve. Instead, we employ more modest values of~$m_i/m_e$, distinguishing between the accretion dynamics of a pair plasma,~$m_i/m_e=1$, and that of a progressively more realistic electron-ion plasma, $m_i/m_e=16 \, \rm and\, 256$. 

The grid spacing in our simulations, parameterized through~$N_r$ and~$N_\theta$, is designed to resolve the plasma microscales, the smallest of which is the Debye length, $\lambda_D = \sqrt{k_B T_0/4\pi n_0 e^2}$. Thanks to our logarithmically stretched radial grid, it suffices to resolve~$\lambda_{\rm D}$ at the outer boundary of the box. The compression of the grid toward inner radii then compensates the plasma compression, keeping the kinetic scales resolved everywhere. We set~$N_r = N_\theta$ in all of our simulations, adjusting~$N_r$ between~$512$ and~$2048$ according to the resolution requirements imposed by the values of~$m_i / m_e$ and~$\sigma_0$ of a given run. In addition, we set the initial magnetic field strength $B_0$ such that the nominal ion gyroradius is $r_L = m_i/eB_0 = r_g$.

%--------------------------------------------------------------------

\section{Accretion of plasma and flare dynamics}\label{sec:Pairs}

\subsection{Global dynamics}
\label{sec:globalphases}
Here we present the behavior of a representative pair-plasma simulation~($m_i = m_e$) with $\sigma_0 =0.05$ and~$N_r = 1024$, illustrated in Fig.~\ref{fig:bigfig}. The top panel of Fig.~\ref{fig:bigfig} shows the magnetic flux threading the black hole, $\Phi_{\rm H}$, the accretion rate, $\dot{M}$, and the bulk electromagnetic dissipation rate, $\dot{\mathcal{E}}$, defined as
\begin{align}
    \Phi_{\rm H} &= \frac{1}{2} \iint  \sqrt{h}\, |B^r|\, \textrm{d}\theta \textrm{d}\phi, 
    \label{eq:phihdef}
    \\
    \dot{M} &= - \iint \sqrt{-g}\, N^r \, \textrm{d}\theta \textrm{d}\phi, \quad \textrm{and}
    \label{eq:mdotdef}
    \\
    \dot{\mathcal{E}} &= \iiint \sqrt{h}\, \textbf{J} \cdot \textbf{E}\, \textrm{d}r \textrm{d}\theta \textrm{d}\phi,
\end{align}
where $\textbf{J}$ is the electric current, $h$ is the determinant of the 3-metric $h_{ij}$ \citep{Komissarov2004}, $g$ is the determinant of the 4-metric $g_{\mu \nu}$, and $N^r$ represents the rest-mass flux averaged over all particle species. Integrals~(\ref{eq:phihdef}) and~(\ref{eq:mdotdef}) are evaluated at the horizon,~$r=r_{\rm H}$. 
The behavior in all of the above quantities is quasi-periodic, with each cycle comprising three distinct phases:
\begin{enumerate}[label=\Roman*.]
\item $\Phi_{\rm H}$ and~$\dot{M}$ both increase linearly in time;
\item The ratio $\Phi_{\rm H}/\sqrt{\dot{M}}$ reaches a saturation value of roughly~$60$, remaining approximately constant as magnetic flux continues to accrue on the horizon;
\item A critical point is reached, and the system undergoes an eruption in which the magnetic flux decreases by a factor of ${\sim}4$ over a timescale of ${\sim} 100 t_g$ -- a very short time compared to the~${\sim}10^3 t_g$ cycle period.
\end{enumerate}
The eruption is accompanied by a sharp rise in $\dot{\mathcal{E}}$, indicating the dissipation of a substantial fraction of the magnetic energy stored during the preceding flux accumulation phases. After the ejection, the cycle restarts, with slow flux accumulation culminating in further abrupt eruptions. All of our simulations exhibit the three phases outlined above, and we refer to these phases -- one, two, and three -- throughout this work.
The first cycles in Fig.~\ref{fig:bigfig} exhibit deviations from the others both in duration and peak magnetic flux storage. This may reflect memory effects associated with the initial conditions. Starting from the fourth cycle, however, the system settles into a more quasi-stationary pattern, with successive cycles exhibiting increasingly uniform behavior.

Four snapshots of the simulation are shown in Fig.~\ref{fig:bigfig}. Three of these snapshots represent the three phases of the simulation cycle; one represents the moment where~$\Phi_{\rm H}$ peaks, from which the eruption (phase three) is triggered. Each snapshot displays spatial maps of the plasma number density $n$, the magnetization $\sigma = B^2/4\pi\rho$, the averaged particle Lorentz factor $\langle \gamma \rangle$, and the electromagnetic energy dissipation $\textbf{J} \cdot \textbf{E}$.
In the spatial maps of Fig.~\ref{fig:bigfig}, the magnetic field divides the box into two topologically distinct zones. One zone is threaded by field lines attached to the black hole. These field lines form a highly magnetized funnel region where plasma accretes by flowing along the magnetic field. The other zone is threaded by field lines not attached to the black hole. These field lines are held in place by a dense and weakly magnetized plasma that builds up along the equator. Unlike in the funnel region, the dense equatorial plasma accretes mostly perpendicular to the field lines, carrying them toward the black hole and, thus, building up the horizon-threading magnetic flux and corresponding funnel zone. As the funnel broadens, oppositely oriented field lines are pushed closer to the equator, building up a radial field reversal. This reversal is supported by an azimuthal electric current that develops in the equatorial plasma.

Starting in phase two, the equatorial current layer thins to the point where intermittent magnetic reconnection events appear, contributing to the regulation of the magnetic flux threading the black hole (second snapshot of Fig.~\ref{fig:bigfig}). These events are too minor in phase two to repulse the heavy plasma channeled in through the equator. However, during phase three, a much more powerful reconnection event (third snapshot of Fig.~\ref{fig:bigfig}) completely arrests and expels the accretion flow (fourth snapshot of Fig.~\ref{fig:bigfig}), preying on the energy stored in the accumulated magnetic field to do so. This energy reservoir depletes once a significant fraction of $\Phi_{\rm H}$ is consumed. At that point, the electromagnetic force can no longer resist the force of gravity, and material infall resumes.

To quantify the energetic efficiency of this process, we characterize key time-averaged values of~$\dot{M}$ and~$\dot{\mathcal{E}}$. We measure:
\begin{align}
    \langle \dot{M} \rangle_{\rm cycle} &= \frac{\int \dot{M} \,\textrm{d}t}{\int \textrm{d}t} = 0.14 \dot{M}_0 \, , \notag \\
    \langle \dot{\mathcal{E}} \rangle_{\rm cycle} &= \frac{\int \Theta \left(\dot{\mathcal{E}} \right) \dot{\mathcal{E}} \,\textrm{d}t}{\int \textrm{d}t} = 0.0044 \dot{M}_0 \, , \quad \mathrm{and} \notag \\
    \langle \dot{\mathcal{E}} \rangle_{\rm flare} &= \frac{\int \Theta \left(\dot{\mathcal{E}} \right) \dot{\mathcal{E}} \,\textrm{d}t}{\int \Theta \left(\dot{\mathcal{E}} \right) \textrm{d}t} = 0.021 \dot{M}_0 \, ,
\end{align}
where, as in Fig.~\ref{fig:bigfig},~$\dot{M}_0 = 4 \pi m_i n_0 r_{\rm inj}^2 v_{\rm th}$.
In the above, integrals are taken over the fourth and fifth accretion-expulsion cycles (last two cycles in the time series of Fig.~\ref{fig:bigfig}) and the Heaviside function,~$\Theta(x)$, selects only positive values of~$\dot{\mathcal{E}}$ (red regions in the~$\dot{\mathcal{E}}$-panel of Fig.~\ref{fig:bigfig}). Positive~$\dot{\mathcal{E}}$ corresponds to electromagnetic energy transferred to particles and thus represents energy made available for emission as observable radiation. Instead, negative~$\dot{\mathcal{E}}$ represents the work of the particles (and ultimately gravity) increasing the electromagnetic energy density within our simulation box.
We can use these measurements to define the black hole efficiency factors
\begin{align}
    \eta_{\rm cycle} \equiv \frac{\langle \dot{\mathcal{E}} \rangle_{\rm cycle}}{\langle \dot{M} \rangle_{\rm cycle}} \simeq 0.03 \quad \mathrm{and} \quad \eta_{\rm flare} \equiv \frac{\langle \dot{\mathcal{E}} \rangle_{\rm flare}}{\langle \dot{M} \rangle_{\rm cycle}} \simeq 0.2 \, .
\end{align}
These efficiency factors are useful in different observational situations. For a dim source, one may wish to integrate over many flare cycles to obtain a significant detection. In that case,~$\eta_{\rm cycle} \langle \dot{M} \rangle_{\rm cycle}$ can be used to estimate the exposure time needed to accumulate a required number of photons. However, for a bright source where observations of single flares are possible,~$\eta_{\rm flare} \langle \dot{M} \rangle_{\rm cycle}$ provides an estimate for the on-flare luminosity. We discuss use cases of both~$\eta_{\rm cycle}$ and~$\eta_{\rm flare}$ in Section~\ref{sec:Discussion}.

\subsection{Particle acceleration in the different phases}

\begin{figure}[h]
    \centering
    \includegraphics[width=\columnwidth]{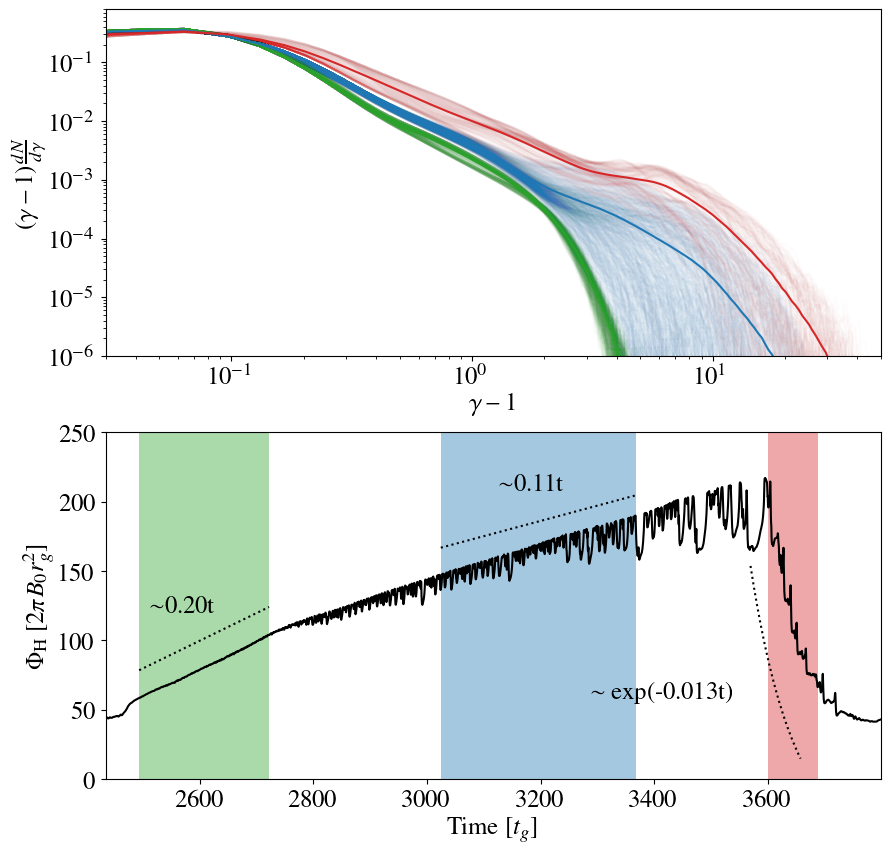}
    \caption{\textbf{Top panel:} Particle energy distributions at different stages of the eruption cycle. Transparent lines show instantaneous distributions; opaque lines represent distributions time-averaged over each phase. Phases one, two, and three are color-coded, respectively, as green, blue, and red.
    \textbf{Bottom panel:} Time evolution of~$\Phi_{\rm H}$ over one eruption cycle. Shaded regions (green, blue, red) indicate the time intervals during which particle distributions are presented in the top panel.}
    \label{fig:spec-phase}
\end{figure}

In addition to regulating accretion, reconnection results in particle acceleration visible in the $\textbf{J}\cdot \textbf{E}$ and $\langle \gamma \rangle - 1$ maps of Fig.~\ref{fig:bigfig}. The intermittent reconnection events of phase two correspond to small flashes of particle acceleration that occur near the horizon, although the accelerated particles are ultimately swallowed by the black hole. During the erupting phase, a substantial fraction of the energy built up in the magnetosphere is abruptly transferred to particles, resulting in much more intense equatorial particle acceleration over a broader range of radii. Some of the accelerated particles escape along the wall of the funnel region (fourth snapshot of Fig.~\ref{fig:bigfig}).

These observations are corroborated by the evolution of the particle energy distribution, shown in Fig.~\ref{fig:spec-phase}. The figure shows instantaneous particle distributions as well as distributions obtained by time-averaging over phases one, two, and three of one cycle of the simulation. All distributions show a thermal bump with a maximum at $\gamma - 1 \sim 0.1 \sim \sqrt{\vartheta_0}$, corresponding to the temperature of injected particles. Particle acceleration manifests as high-energy nonthermal tails in the distributions.

In phase one (green), the particle spectrum appears rather soft with little variability. The absence of magnetic reconnection from this phase results in little to no particle acceleration. In contrast, during the reconnection-regulated phase (blue), intermittent reconnection accelerates particles episodically, resulting in a somewhat extended and highly variable tail in the distributions. The erupting phase (red) shows the strongest particle acceleration, with more particles being accelerated overall and with a higher maximum particle energy. A large number of magnetic reconnection studies in 2D slab geometry show that, in pair plasmas, the particle energy distribution cuts off (or at least turns over) around~$\gamma \sim \sigma$, corresponding to an approximately equal partitioning of the available magnetic energy density among the accelerated particles:~$\gamma n m_e c^2 \sim B^2 / 8 \pi$ (see \citealt{sironi_etal_2025} and references therein). During the eruption phase, the particle distribution turns over at around~$\gamma \sim 10$, hinting that the magnetization witnessed by particles in the reconnection layer is~$\sigma \sim 10$, consistent with the values in the funnel region shown in Fig.~\ref{fig:bigfig}.

\subsection{Robustness of the dynamics}
\label{sec:robustness}

All of the above phases and particle acceleration results (with the exception of the separation between species that develops for~$m_i > m_e$, described later on) hold for all of the simulations presented in this work. However, we also conducted a broad suite of supplementary test simulations, pushing the numerical parameters beyond the confines of those presented here. Below, we comment on the robustness of the results described so far in light of these test campaigns.
\begin{itemize}
    \item Effect of~$\sigma_0$:
    
    We explored initial magnetizations ranging from $\sigma_0 = 0.3$ to $\sigma_0 = 0.03$. We observed that the magnetization on the event horizon, $\sigma_{\rm H}$, reaches ${\sim}10$ in all cases. In addition, the maximum value of $\Phi_{\rm H}$ (reached immediately before eruption) decreases with increasing $\sigma_0$. Less magnetic flux accumulation is required to reach~$\sigma_{\rm H}=10$ for a higher~$\sigma_0$. However, if the simulation box is not big enough,~$\sigma_{\rm H}$ may not reach~$10$, as described below.
    \\
    \item Effect of box size:
    
    Reducing $r_{\rm max}$ by $2/3$ resulted in a decreased maximum value of $\Phi_{\rm H}$ with respect to the simulation presented above, with $\sigma_{\rm H}$ no longer saturating near~$10$. On the contrary, increasing $r_{\rm max}$ by a factor of~$5/3$ altered neither the maximum~$\Phi_{\rm H}$ nor the saturated~$\sigma_{\rm H}$ value. We interpret this as meaning that our runs are converged, possessing enough initial magnetic flux in the box to supply the flux~$\Phi_{\rm H}$ necessary to achieve a horizon magnetization of~$\sigma_{\rm H} \sim 10$.
    \\
    \item Effect of initial temperature:

    We did not alter $\vartheta_0 = k_{\rm B} T_0 / m_i$ much, keeping it close to the Bondi limit,~$2r_g/r_{\rm pml}$ (see Sect.~\ref{sec:Methods}). However, we noticed that plasma instabilities such as mirroring arose in the funnel region for simulations with $\sigma < 0.1$ and $\beta_{\rm plasma} > 1$ (a regime barely accessed by our runs). This feature has also been observed in the context of black hole accretion by \cite{2023PhRvL.130k5201G}.
\end{itemize}

In addition, we note that our simulations consider an unrealistically small Bondi radius, which lies at the edge of our simulation domain ($\sim 30 r_g$). Modeling realistic Bondi accretion is not computationally accessible with kinetic simulations. However, many of the essential features we observe, including magnetic flux eruptions, have also been witnessed in GRMHD simulations with a realistic material supply from larger scales \citep{2021MNRAS.504.6076R}. In light of this broad agreement with prior GRMHD work, and also thanks to the robustness of our results to variations in numerical parameters discussed here, we believe that the reduced-scale dynamics observed in our kinetic simulations may scale up to larger systems.

\section{Effect of scale separation between the species}\label{sec:Ions}

\subsection{Scale separation features}

\begin{figure}[h]
    \centering
    \includegraphics[width=\columnwidth]{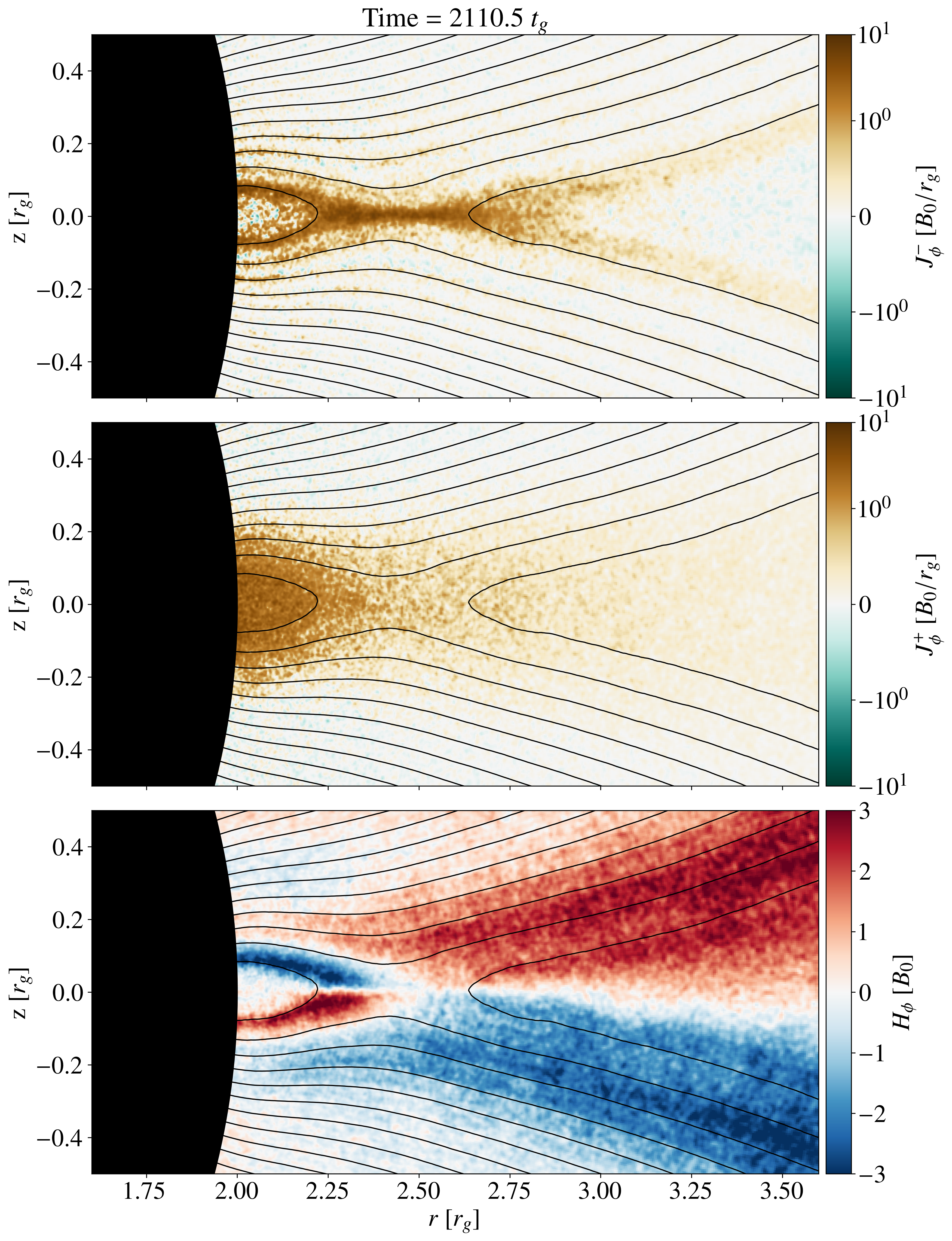}
    \caption{\textbf{Top:} map of the azimuthal electric current carried by the electrons ($J^-_\phi$). \textbf{Center:} map of the azimuthal electric current carried by the ions ($J^+_\phi$). \textbf{Bottom:} map of the auxiliary azimuthal magnetic field $H_\phi$. Magnetic field lines are represented in solid black lines for all panels. The mass ratio is $m_i/m_e = 256$ in the depicted simulation.}
    \label{fig:J-Hphi_ions}
\end{figure}

In this second part of our work, we study the effect of the mass ratio, $m_i/m_e$, on the dynamics. We present simulations with $m_i/m_e \in [1, 16, 256]$, increasing the initial magnetization to $\sigma_0 =0.3$ to make this range numerically feasible. All cases exhibit the same three-phase cyclic dynamics as detailed for pair plasmas in Section~\ref{sec:Pairs}. The typical eruption timescale is still~${\sim} 100 t_g$ and~$\sigma_{\rm H}$ still peaks at~${\sim}10$. However, due to the increased~$\sigma_0$ (see Section~\ref{sec:robustness}), the maximum value of $\Phi_{\rm H}$ is somewhat reduced, reaching~${\sim}40{-}50$. Overall, the marked similarity to the pair-plasma case suggests that pair plasmas may be used to model eruptive magnetospheric dynamics at substantially lower computational cost.

However, a nonzero inter-species scale separation does impact certain microscopic details of the evolution. First, due to the higher ion inertia, ions show a somewhat different distribution of equatorial current than electrons. As shown in the top panels of Fig.~\ref{fig:J-Hphi_ions}, the azimuthal current carried by the electrons is more intense and organized on smaller spatial scales, whereas the ion current is more diffuse.
The electron-ion scale separation is also highlighted at reconnection X-points. Electrons approach X-points more closely than ions do before becoming demagnetized and crossing into the outflow region. This decoupling between species generates a poloidal electric current in the outflow from X-points, inducing a quadrupolar magnetic field $H_\phi$ \citep{werner_etal_2018}, as shown in the bottom panel of Fig.~\ref{fig:J-Hphi_ions}.

\subsection{Impact on particle acceleration}
\label{sec:ions-accel}

\begin{figure}[h]
    \centering
    \includegraphics[width=\columnwidth]{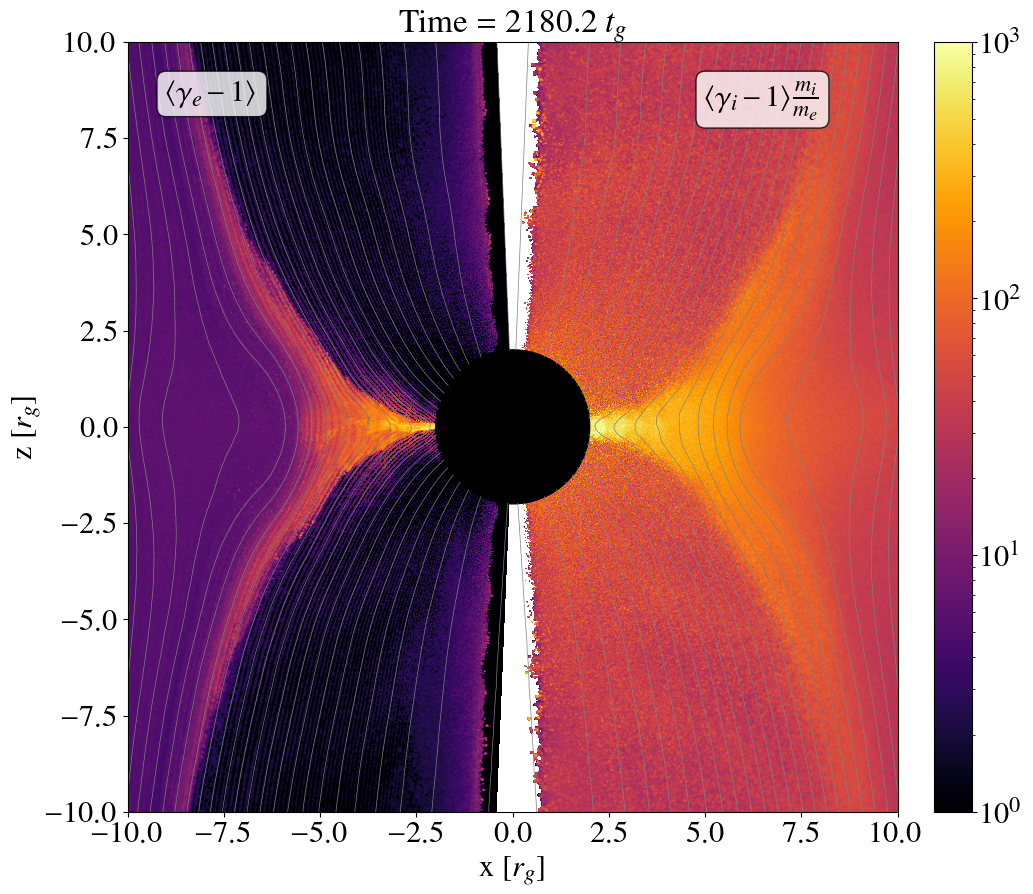}
    \caption{Map of the local average particle energy for electrons (left) and ions (right) for the simulation with $m_i/m_e=256$. The energies are normalized to $m_e c^2$ and grey lines represent magnetic field lines.}
    \label{fig:map_ene}
\end{figure}

Perhaps the starkest difference between electrons and ions is with respect to particle energization. As previously explained, magnetic reconnection is the key mechanism that accelerates particles during flux eruptions. If the liberated energy were divided evenly among all particles, then the electrons would reach much higher Lorentz factors than the ions. Even though the species do not share energy precisely equally in reality \citep{werner_etal_2018, werner_uzdensky_2024, comisso_2024}, this is a small effect compared to the mass ratio in deciding the characteristic Lorentz factors that particles attain. For example, Fig.~\ref{fig:map_ene} displays a map of the average kinetic energy for each species during a magnetic flux eruption. The strong discrepancy between electron and ion energies in the funnel region is merely an initialization artifact; electrons are injected colder than ions (Section~\ref{sec:Methods}). Despite the initial temperature gap, the species achieve comparable average energies once processed by reconnection, indicating characteristic Lorentz factors of accelerated electrons a factor of~$m_i/m_e$ higher than those of accelerated ions.

Here again, we see that the spatial distribution of accelerated ions is more diffuse than that of high-energy electrons. High-energy electrons reside both in the equatorial reconnection layer and along the wall of the funnel region. This suggests both locations as possible sites of high-energy emission during flux eruption events. 

\begin{figure}
    \centering
    \includegraphics[width=\columnwidth]{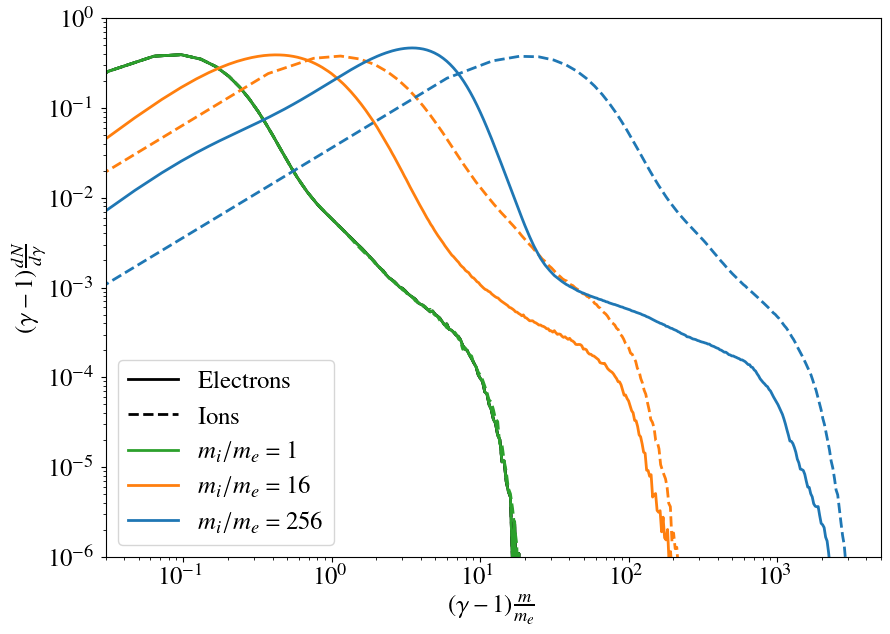}
    \caption{Time-averaged energy distributions of the particles during the erupting phase for all electron-ion simulations (including the reference $m_i = m_e$ run). Ions are represented as dashed lines and electrons as solid lines.}
    \label{fig:spec_ions}
\end{figure}

We also compare the particle energy distributions of the species as a function of mass ratio in Fig.~\ref{fig:spec_ions}. The figure shows snapshots taken during the third erupting phase for all the mass ratios considered. The ion energy distributions remain virtually identical when increasing the mass ratio; the only difference between ion distributions is a horizontal shift consistent with their higher rest-mass energy. This again demonstrates the robustness of the overall dynamics with respect to the mass ratio: the species dominating the plasma inertia behaves similarly in all cases.

However, qualitative differences emerge for the electrons as~$m_i/m_e$ is increased. The extent of the nonthermal tail of the electron distribution is enhanced, covering almost two decades for $m_i/m_e=256$ instead of barely one for $m_i=m_e$. The power-law slope also appears to harden for higher mass ratios, reaching an index of $d \log N / d \log \gamma \sim -1.7$ at the highest mass ratio. We attribute both effects to the increased magnetization felt by the electrons near the event horizon, $\sigma_{\mathrm{H},e} \sim (m_i/m_e)\sigma_{\rm H} \sim 10^3$. With more magnetic energy available per unit of rest-mass energy, the tail of the electron distribution turns over at a higher Lorentz factor~${\sim}\sigma_{\mathrm{H},e}$. The hardening of the power-law slope with increasing magnetization probably owes to a similar mechanism as has been previously observed in 2D simulations of pair-plasma \citep{guo_etal_2014, werner_etal_2016} and electron-ion plasma \citep{werner_etal_2018} reconnection in slab geometry.

This robust trend towards stronger nonthermal electron acceleration with increasing mass ratio can be extrapolated to the full proton-electron value, $m_i/m_e = 1836$. We expect that, in this case, the electron energy distribution should turn over or cut off at $\gamma \sim \sigma_{\mathrm{H},e} \sim (m_i/m_e) \sigma_{\rm H} \sim 10^4$. However, we expect the power-law slope to plateau in the ultrarelativistic limit \citep{guo_etal_2014, werner_etal_2016, werner_etal_2018, ball_etal_2018} and, thus, to remain roughly constant between~$m_i/m_e = 256$ (corresponding to~$\sigma_{\mathrm{H},e} \sim 10^3$) and~$m_i/m_e = 1836$ (for which~$\sigma_{\mathrm{H},e} \sim 10^4$).

%--------------------------------------------------------------------

\section{Modeling the cyclic activity}\label{sec:Model}

In what follows, we present several theoretical aspects of the three-phase cycle outlined in Section~\ref{sec:globalphases}. We discuss major physics points of phases one, two, and three in Sections~\ref{sec:phase1},~\ref{sec:phase2}, and~\ref{sec:phase3}, respectively. Then, in Section~\ref{sec:transitions}, we argue that the transitions between phases are governed by the progressive thinning of the equatorial current layer.

\subsection{Phase I: Ideal accretion}
\label{sec:phase1}
This is the first of two phases where the plasma accretes onto the black hole, dragging magnetic field lines with it. The hallmark of this phase is that it respects the ideal flux-freezing condition of magnetohydrodynamics (MHD).
The accretion of flux is thus tied to that of matter, as we demonstrate below.

To show how material and magnetic flux transport are linked, we here derive and solve the magnetic flux transport equation. 
To arrive at this equation, we need three ingredients: (i) the ideal MHD condition, which can be expressed as
\begin{align}
    \textbf{E} + \textbf{V} \times \textbf{B} = 0 \,
    \label{eq:fluxfreezing}
\end{align}
in the~$3{+}1$ formalism of \citet{Komissarov2004}, where~$\mathbf{V}$ is the bulk fluid 3-velocity;~(ii) the magnetic flux threading a spherical cap of radius~$r$ and half opening angle~$\theta$,
\begin{align}
    \Phi(r,\theta,t) = 2 \pi \int_0^{\theta} \sqrt{h} \, B^r \mathrm{d}\theta \, ;
    \label{eq:fluxfunction}
\end{align}
and~(iii) the Maxwell-Faraday law
\begin{align}
    \partial_t \textbf{B} = - \nabla \times \textbf{E} \, .
    \label{eq:faraday}
\end{align}
Integrating Eq.~(\ref{eq:faraday}) over the same spherical cap as Eq.~(\ref{eq:fluxfunction}) yields the time-evolution equation for~$\Phi$:
\begin{align}
    \partial_t \Phi = -2 \pi \int_0^{\theta} \partial_\theta E_\phi \, \mathrm{d} \theta = -2 \pi E_\phi \, .
    \label{eq:dphidt}
\end{align}
Finally, plugging in the~$\phi$-component of Eq.~(\ref{eq:fluxfreezing}),
\begin{align}
    E_\phi + \sqrt{h} \left(V^r B^\theta - V^\theta B^r \right) = 0 \, ,
\end{align}
and using
\begin{align}
    2\pi \sqrt{h} B^r = \partial_\theta \Phi \quad \mathrm{and} \quad 2\pi \sqrt{h} B^\theta = -\partial_r \Phi \, ,
\end{align}
gives the equation for magnetic flux transport,
\begin{align}
    \partial_t \Phi = 2 \pi \sqrt{h} \left( V^r B^\theta - V^\theta B^r \right) = -V^r \partial_r \Phi - V^\theta \partial_\theta \Phi \, .
\end{align}
We now specialize to the equatorial plane,~$\theta=\pi/2$, for which the top-down symmetry of our problem dictates~$V^\theta = 0$. This results in the simplified flux-advection equation
\begin{align}
    \left( \partial_t + V^r \partial_r \right)\Phi = 0 \, .
    \label{eq:phiadvect}
\end{align}

The equatorial material accretion velocity,~$V^r$, dictates magnetic flux deposition onto the black hole. Conversely, if the flux transport,~$\Phi(r,\pi/2,t)$, is known, then the equatorial accretion velocity follows. Given that we have measured empirically that~$\Phi(r_{\rm H},\pi/2,t) = \Phi_{\rm H}$ grows linearly in time, let us ask the following question: What kind of velocity profiles,~$V^r$, are consistent with~$\dot{\Phi}_{\rm H} = \, \mathrm{constant}$? We consider self-similar, time-independent profiles~$V^r = -\mathcal{C}/r^\xi$, with proportionality constant~$\mathcal{C} > 0$. Then, we solve Eq.~(\ref{eq:phiadvect}) by the method of characteristics, searching for the~$\xi$ that reproduces~$\partial_t \Phi_{\rm H} = \mathrm{const.}$

\begin{figure}
    \centering
    \includegraphics[width=\columnwidth]{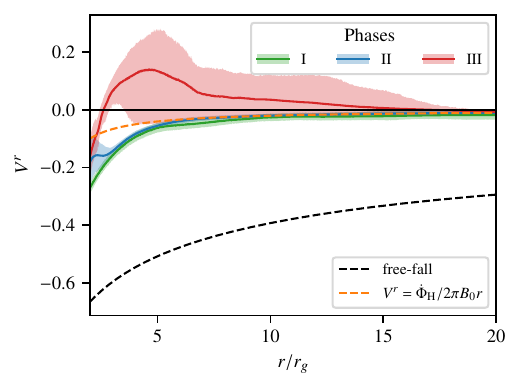}
    \caption{Profiles of~$V^r$ measured from the simulation presented in Section~\ref{sec:Pairs}. Solid lines denote time-averages over the intervals defined in Fig.~\ref{fig:spec-phase}; envelopes denote one-sigma percentiles for each averaging interval. The orange dashed curved corresponds to Eq.~(\ref{eq:vrsoln}) where~$\dot{\Phi}_{\rm H}$ is measured during phase one on Fig.~\ref{fig:spec-phase}. The dashed black line denotes the free-fall velocity of a particle starting from rest at infinity (Eq.\ref{eq:vrff}).}
    \label{fig:vr}
\end{figure}

For the ansatz,~$V^r=-\mathcal{C}/r^\xi$, the solution to Eq.~(\ref{eq:phiadvect}) is
\begin{align}
    \Phi(r,\pi/2,t) = \Phi(r_0,\pi/2,0) = \pi B_0 r_0^2 \, ,
\end{align}
where
\begin{align}
    r_0 = \left[ \mathcal{C} \left(1 + \xi \right) t + r^{1+\xi} \right]^{1/(1+\xi)} \, ,
\end{align}
and we have used our simulation initial condition,~$\Phi(r_0,\pi/2,0)=\pi B_0 r_0^2$. We search to match the solution at the horizon,~$r=r_{\rm H}$, to a linearly growing function of~$t$. For~$r=r_{\rm H}$ and~$t\gg r_{\rm H}^{1+\xi}/[\mathcal{C} (1+\xi)]$, we have
\begin{align}
    \Phi_{\rm H} = \Phi(r_{\rm H},\pi/2,t) \simeq \pi B_0 \left[ \mathcal{C} (1+\xi) t \right]^{2/(1+\xi)} \, .
\end{align}
Thus, for a self-similar, time-independent velocity profile~$V^r$, the flux threading the horizon,~$\Phi_{\rm H}$, grows linearly in time if and only if~$\xi = 1$. Rewriting the constant~$\mathcal{C}$ in terms of~$\dot{\Phi}_{\rm H}$ then gives
\begin{align}
    V^r = - \frac{\dot{\Phi}_{\rm H}}{2 \pi B_0 r} \, .
    \label{eq:vrsoln}
\end{align}
Next, we examine how well the profile (Eq.~\ref{eq:vrsoln}) matches the equatorial accretion velocity of our simulations. We present in Fig.~\ref{fig:vr} the equatorial profiles of~$V^r$ measured from our simulation presented in Section~\ref{sec:Pairs}. To compare with Eq.~(\ref{eq:vrsoln}), we plug in~$\dot{\Phi}_{\rm H} \simeq 0.20$, measured during phase one in Fig.~\ref{fig:spec-phase}.

The measured~$V^r$ profiles agree very well with the theoretical prediction~(Eq.~\ref{eq:vrsoln}) far from the black hole. Close to the black hole, there is some deviation, suggesting that some amount of magnetic diffusivity may be present near the horizon even before reconnection starts to act. Importantly, however, the accretion velocities are all much slower than that of a particle that freely falls from rest at infinity,
\begin{align}
    V^{r}_{\rm ff} = - \left(\frac{r}{r_{\rm H}}\right)^{-1/2} \frac{1-(r/r_{\rm H})^{-1}}{1-\left( r/r_{\rm H} \right)^{-3/2}} \, ,
    \label{eq:vrff}
\end{align}
indicating that magnetic tension is strong enough to significantly slow gravitational infall even in phase one.

\subsection{Phase II: Reconnection-regulated accretion}
\label{sec:phase2}
Matter and magnetic fields continue to accrete in this phase. The crucial difference from the preceding phase is the appearance of intermittent near-horizon reconnection events. These are traced by the high-frequency noise in the time series of~$\Phi_{\rm H}$ in this phase~(Fig.~\ref{fig:spec-phase}).
Because reconnection allows particles to locally slip across magnetic field lines, less flux accrues on the black hole per unit of mass accreted. In fact, the average equatorial accretion velocity~$V^r$, as shown in Fig.~\ref{fig:vr}, is not that different in this phase from phase one. Nevertheless, reconnection-mediated magnetic diffusion reduces the average growth rate of~$\Phi_{\rm H}$ by a factor of~$\simeq 2$ with respect to the first phase, as shown in Fig.~\ref{fig:spec-phase}.

Another distinguishing aspect of this phase is that the magnetic field near the horizon strongly regulates accretion. Namely, the magnetic tension across the equator approximately matches the force of gravity on the heavy equatorial plasma. This fixes the ratio of~$\Phi_{\rm H} / \sqrt{\dot{M}}$ to roughly~$60$, as we derive below.

In the~$3{+}1$ formalism of \citet{Komissarov2004}, FIDOs measure a gravitational pull on the plasma of
\begin{equation}
    f^G_r = - \rho \frac{\partial_r \alpha}{\alpha},
\end{equation}
where $\alpha = 1/\sqrt{1+2r_g/r}$ is the lapse function,~$\rho$ is the FIDO-measured mass density, and we assume a nonrelativistic plasma. The expression for the Lorentz force measured by FIDOs is 
\begin{equation}
    f^L_r = - \sqrt{h} \frac{J^\phi}{\alpha} B^\theta,
    \label{eq:flfido}
\end{equation}
where
\begin{equation}
    J^\phi = \frac{1}{4\pi}(\nabla \times H)^\phi = - \frac{1}{4\pi\sqrt{h}} \partial_\theta H_r.
    \label{eq:ampere}
\end{equation}
The auxiliary magnetic field,~$H_r$, is related to $B_r$ by $H_r = \alpha g_{rr}B^r = B^r/\alpha$. In the vicinity of an X-point,~$B^r$ switches sign from $B^r_{\rm up}$ to $-B^r_{\rm up}$ over an angular width $\Delta \theta = 2 \delta/r_X$. Here,~$r_X$ is the radial position of the X-point and~$\delta$ is the local half-thickness of the current sheet. Plugging this into Eq.~(\ref{eq:ampere}) gives
\begin{equation}
    J^\phi = \frac{1}{4\pi\alpha\sqrt{h}} \frac{r_X B^r_{\rm up}}{\delta}.
    \label{eq:jphuestimate}
\end{equation}
At the X-point, a vertical magnetic field component $B^\theta_X$ is created by the reconnection of the upstream magnetic field $B^r_{\rm up}$. In the orthonormal basis ($B^{\hat{i}}=\sqrt{g_{ii}} B^i$), we expect that the reconnection rate,~$\beta_{\rm rec}$, relates the $r$ and $\theta$ components of the magnetic field by
\begin{equation}
    \beta_{\rm rec} \sim \frac{B^{\hat{\theta}}_X}{B^{\hat{r}}_{\rm up}} = \sqrt{\frac{g_{\theta \theta}}{g_{rr}}} \frac{B^\theta_X}{B^{r}_{\rm up}} = r_X \alpha \frac{B^\theta_X}{B^{r}_{\rm up}},
    \label{eq:betarecfieldratio}
\end{equation}
Using Eqs.~(\ref{eq:flfido}),~ (\ref{eq:jphuestimate}), and~(\ref{eq:betarecfieldratio}), we recover the Lorentz force estimate,
\begin{equation}
    f^L_r \sim  \frac{\beta_{\rm rec}}{4\pi \alpha^3 \delta} \left(B^r_{\rm up}\right)^2.
\end{equation}
Balancing the gravitational and Lorentz forces then gives
\begin{equation}
    \frac{\beta_{\rm rec}}{4\pi \alpha^3 \delta} \left(B^r_{\rm up}\right)^2 \sim \rho \frac{\partial_r \alpha}{\alpha}.
\end{equation}
We approximate~$B^r_{\rm up} \sim \alpha \Phi_{\rm H}/2\pi r^2_X$ and assume that a significant part of the accretion flow passes through the current layer (which we verified in our simulations) so that $\dot{M}(r=r_{\rm H})\sim \dot{M}(r=r_X) \sim \rho V_{\rm acc} 4\pi r_X\delta / \alpha$. In this manner, we replace~$B^r$ with~$\Phi_{\rm H}$ and~$\rho$ with~$\dot{M}$ to get
\begin{equation}
    \frac{ V_{\rm acc} \beta_{\rm rec}}{\alpha} \frac{\Phi^2_{\rm H}}{4\pi^2 r^3_X} = \dot{M}\partial_r{\alpha}.
\end{equation}
Rearranging, we arrive at
\begin{equation}
    \frac{\Phi_{\rm H}}{\sqrt{\dot{M}}} = 2 \pi r^{3/2}_X \sqrt{\frac{\alpha\partial_r \alpha}{V_{\rm acc} \beta_{\rm rec}}}.
\end{equation}
During phase two, the X-point location is observed in the simulations to lie on average at $r_X\sim3 r_g$, and the accretion velocity is measured to be $V_{\rm acc} \sim 0.1$. Plugging these in and assuming $\beta_{\rm rec} \sim 0.1$, we find that
\begin{equation}
    \frac{\Phi_{\rm H}}{\sqrt{\dot{M}}} \sim 60.
\end{equation}
This derivation was made using strong assumptions on the dynamics and the accretion flow, and should be explored in the context of more realistic accretion conditions. However, the fact that our simplified setup of accretion and this analytical scaling match the larger-scale GRMHD estimates (e.g. \citealt{2002ApJ...566..137I, 2021MNRAS.504.6076R}) indicates that it might be robust upon more realistic accretion flow conditions.

Conversely, the above arguments cannot predict $\sigma_{\rm H}$. The latter depends on the mass density within the funnel, whereas the model in this section constrains only the mass density near the equator. We expect to conduct a more comprehensive theoretical and numerical exploration of $\sigma_{\rm H}$ in future work and, in particular, to examine its dependence on the black hole spin.

\subsection{Phase III: Flux eruption}
\label{sec:phase3}
The delicate balance between magnetic tension and gravity that characterizes the second phase sets the stage for the eruptive phase. This is because reconnection tends to convert horizontal field~($B^r$) into vertical field~($B^\theta$). Thus, when a sudden, strong reconnection event occurs, it rapidly builds up~$B^\theta$, disrupting the balance between magnetic tension and gravity and triggering an eruption.
During the eruption, a macroscopic reconnection zone forms at the equator, removing flux from the black hole and expelling it outward. The rate at which magnetic flux decays off of the horizon is governed by the collisionless relativistic reconnection rate,~$\beta_{\rm rec} \sim 0.1$, and follows an exponential decay law (Fig.~\ref{fig:spec-phase}). Following models developed by \cite{Crinquand2021} and \cite{Bransgrove2021} we derive this decay law below.

FIDOs just upstream of the reconnection layer (i.e., slightly above or below the equator) measure a ratio of the electric-to-magnetic field equal to the reconnection rate:
\begin{equation}
    \frac{D^{\hat{\phi}}}{B^{\hat{r}}} \sim \beta_{\rm rec},
\end{equation}
where $D^{\hat{i}} = \sqrt{g_{ii}} D^i$ and $\beta_{\rm rec}$ is the reconnection rate expectation in flat spacetime. Assuming a split monopolar magnetic geometry near the event horizon, and placing the FIDO above the X-point, we have
\begin{equation}
    D^\phi = \beta_{\rm rec} \sqrt{\frac{g_{rr}}{g_{\phi\phi}}} B^r = \beta_{\rm rec} \sqrt{\frac{g_{rr}}{g_{\phi\phi}}} \frac{\alpha\Phi_{\rm H}}{2\pi r_X^2} \, .
\end{equation}
Neglecting any $B^\theta$, $E_\phi$ is related to $D^\phi$ by
\begin{equation}
    E_\phi \sim \alpha g_{\phi\phi} D^\phi. 
\end{equation}
Given that $\partial_t \Phi = - 2 \pi E_\phi$~(Eq.~\ref{eq:dphidt}), $g_{rr} = 1/\alpha^2$, and $g_{\phi \phi} = r^2 \sin^2{\theta}$, we get
\begin{equation}
    \partial_t \Phi = - \beta_{\rm rec}\frac{\alpha \sin \theta}{r_X} \Phi_{\rm H} \, .
\end{equation}
We assume that flux loss is transferred from $r_X$ to $r_{\rm H}$ much faster than it reconnects at $r_X$, such that we may write~$\Phi(r_X,\pi/2)\simeq\Phi(r_{\rm H},\pi/2)=\Phi_{\rm H}$. This gives,
\begin{equation}\label{eq:flux_phase3}
    \frac{\mathrm{d} \Phi_{\rm H}}{\mathrm{d}t} = - \frac{\alpha\beta_{\rm rec}}{r_X} \Phi_{\rm H} \, ,
\end{equation}
an exponential decay law that matches the behavior in Fig.~\ref{fig:spec-phase}. In that figure, we fit a decay rate of~$\alpha \beta_{\rm rec}/r_X\sim 0.013 t^{-1}_g$, which implies, for $r_X \sim 3 r_g$, that $\beta_{\rm rec}\sim 0.05$. This is a factor of two lower than the expectation for relativistic magnetic reconnection in flat spacetime. We emphasize, however, that this estimate of~$\beta_{\rm rec}$ should be seen as an average over the whole eruption, during which the upstream plasma properties (such as the magnetization) change. We believe that the evolving upstream, as well as the spherical geometry and the presence of a vertical magnetic field, may account for the slightly slower~$\beta_{\rm rec}$ measured here than in typical slab-geometry simulations.

\subsection{Phase transitions dictated by current layer stability}
\label{sec:transitions}
 
\begin{figure}[h]
    \centering
    \includegraphics[width=\columnwidth]{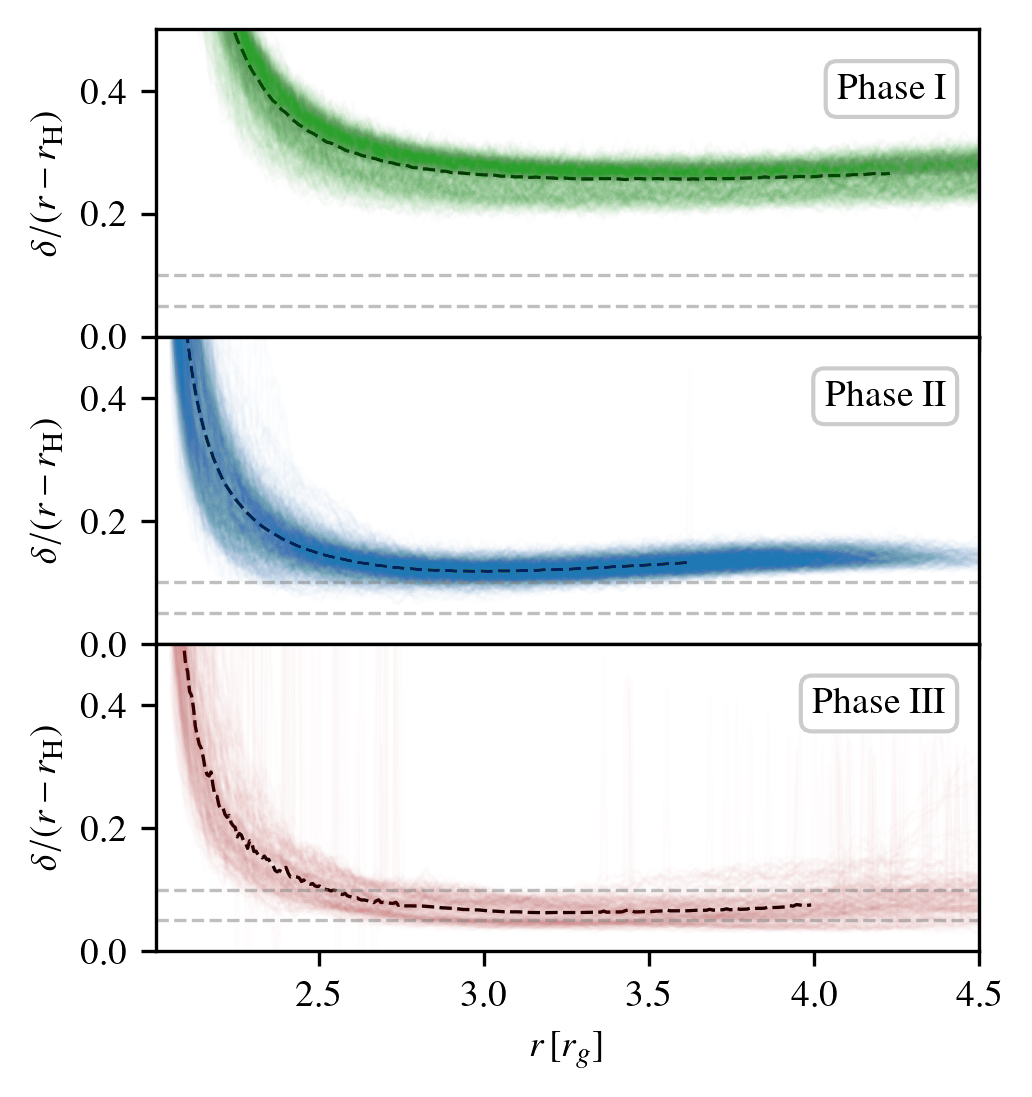}
    \caption{Traces of~$\delta(r) / (r-\rh)$ at specific times within (transparent lines), and time-averaged over (solid dashed lines), each phase.}
    \label{fig:cswidth}
\end{figure}

Sections~\ref{sec:phase1}-\ref{sec:phase3} underscore the role of reconnection in distinguishing each phase.
As the layer thins, it becomes progressively more prone to tearing \citep{zelenyi_krasnoselskikh_1979, zenitani_hoshino_2008}, the formation of X-points and plasmoids, and, thus, to the rapid reconnection of magnetic flux \citep{shibata_tanuma_2001, loureiro_etal_2007, uzdensky_etal_2010}. We argue in this section that the transitions between phases correspond to critical tearing instability thresholds of the current sheet.

To begin with, we measure the half-thickness,~$\delta$, of the equatorial current layer as a function of radius,~$r$, and time,~$t$. These measurements are presented in Fig.~\ref{fig:cswidth}. In the figure, we normalize~$\delta(r)$ by~$r-r_{\rm H}$, the latter being a proxy for the length of the current sheet. Figure~\ref{fig:cswidth} suggests that phase one corresponds to a half opening angle,~$\delta(r) / (r - r_{\rm H})$, greater than~$0.1$. In phases two and three, the half-opening angle is roughly~$0.1$ and~$0.05$, respectively. 

These critical opening angles correspond to tearing instability thresholds of the equatorial current sheet. The rate of collisionless plasmoid-mediated relativistic reconnection is~$\beta_{\rm rec} \simeq 0.1$. This corresponds to the aspect ratio of elementary current layers between small-scale plasmoids. On average, once the distance between two plasmoids exceeds~$\beta_{\rm rec}^{-1} \simeq 10$ times the width of the current sheet between them, that current sheet tears and spins off another plasmoid \citep{shibata_tanuma_2001, loureiro_etal_2005, loureiro_etal_2007, uzdensky_etal_2010, cerutti_etal_2014}.

When~$\delta/(r-r_{\rm H}) = 0.1$, the magnetic field above and below the equatorial current sheet opens at the same angle as the separatrix field lines near a single reconnection X-point. This is conducive to the formation of single X-points, which sporadically appear just beyond the horizon in this regime. They are, however, short-lived, falling quickly into the black hole. Thus, in phase two, the current sheet episodically undergoes single-X-point reconnection \citep{ji_daughton_2011}, with the X-point formation site being forced by the global field geometry to lie close to the event horizon.

\begin{figure*}[h!]
    \centering
    \includegraphics[width=\textwidth]{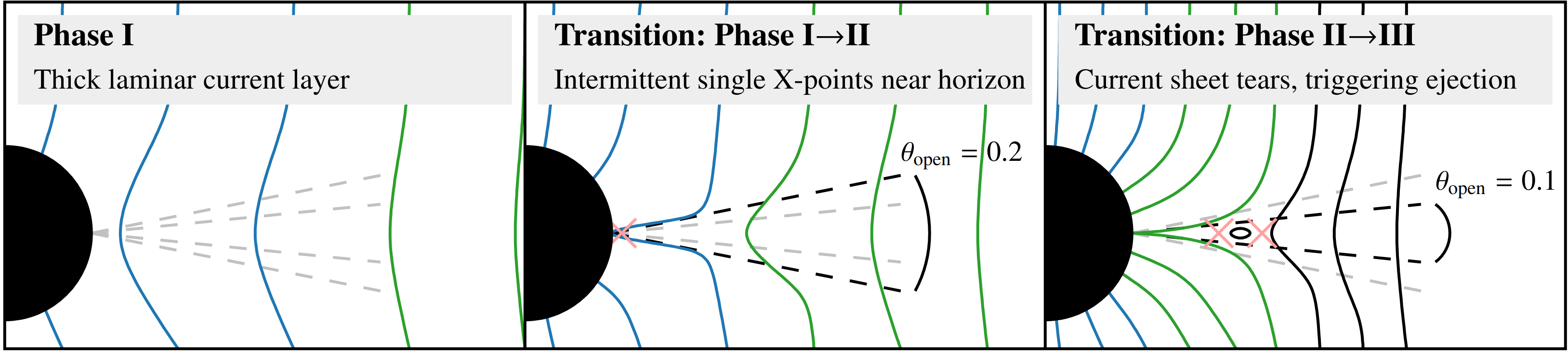}
    \caption{Conceptual three-phase accretion cycle. Transitions between phases coincide with critical opening angles of the magnetic field lines about the equator. Blue and green field lines are accreted during phases one and two, respectively; black field lines never reach the black hole.}
    \label{fig:phase_diagram}
\end{figure*}

Once~$\delta/(r-r_{\rm H})$ decreases down to~$0.05$, a new, multiple X-point regime is attained. This is because the equatorial current layer is now so thin that its \textit{full} width,~$2\delta$, becomes~$10$ times smaller than its \textit{full} length,~$r-r_{\rm H}$. The layer thus becomes statistically likely to form X-points far from the horizon, developing a rapidly reconnecting plasmoid chain \citep{shibata_tanuma_2001, uzdensky_etal_2010, ji_daughton_2011}.
Whereas the exhaust from a single X-point is characterized by a relatively small vertical magnetic field,~$B^{\hat{\theta}} \sim \beta_{\rm rec} B^{\hat{r}}$, in this new fully nonlinear stage of the tearing instability, characterized by circular plasmoids, the vertical field becomes comparable to the upstream, unreconnected field:~$B^{\hat{\theta}} \sim B^{\hat{r}}$. This amplification upsets the equilibrium between magnetic tension and gravity that previously held during phase two, thus triggering the eruption phase.

These ideas are summarized in Fig.~\ref{fig:phase_diagram}. The transition from phase one to phase two corresponds to the moment when the half-opening angle at the horizon reaches~$\theta_{\rm open} \simeq \beta_{\rm rec} = 0.1$, triggering single X-point reconnection. The subsequent transition to phase three occurs once the layer further thins to the point,~$\theta_{\rm open} = \beta_{\rm rec} / 2 = 0.05$, marking the transition to the multiple X-point, plasmoid-dominated reconnection regime.

\section{Astrophysical applications}\label{sec:Discussion}
Generally, we only expect strong nonthermal radiation signatures from the third (eruptive) phase of each flux accumulation cycle. Reconnection does not act in the first phase, and so nonthermal particle acceleration and concomitant nonthermal radiation should be virtually absent. We also expect the second phase to produce very little high-energy emission. Although intermittent reconnection occurs in this phase, it happens very close to the horizon with the emitting particles primarily on infalling trajectories. This presents unfavorable conditions for high-energy radiation to escape the magnetosphere. In contrast to these first two phases, the erupting phase exhibits the strongest reconnection, with a significant fraction of the accelerated particles lying farther away from the black hole and moving outward. We therefore expect most of the observable high-energy nonthermal emission to arise during this phase.

Next, we outline specific astrophysical implications of our results with respect to two different types of low-luminosity black holes. We focus primarily on infrared (IR) and X-ray flares observed from \sgr. Secondarily, we speculate on radiation that may be produced by a mechanism similar to our model from isolated stellar-mass black holes interacting with the Galactic interstellar medium (ISM).

\subsection{\sgr{} IR and X-ray flares}
\label{sec:sgrA}
We organize our discussion around the following observed characteristics of IR and X-ray flaring in \sgr:
\begin{itemize}
    \item Duration:
    
    The eruption phase lasts ${\sim} 100 t_g$ in all our simulations -- a robust timescale tied to the universal rate at which magnetic reconnection removes flux from the black hole (Section~\ref{sec:phase3}). Adopting the mass of \sgr,~${\sim}4\times10^6 M_\odot$, this corresponds to a flare duration of ${\sim}30$ min, broadly consistent with the timescales of the X-ray and IR \sgr{} flares \citep{genzel_etal_2003, ghez_etal_2004, porquet_etal_2008, gravity_etal_2021, vonfellenberg_etal_2025}.
    \\
    \item Recurrence timescale:
    
    Our simulations show a cycle period of about~$10^3 t_g$. Recently, \citet{jacquemin-ide_etal_2025} argued, in the context of GRMHD simulations, that this timescale arises from the balance of magnetic flux advection and diffusion near the black hole, a balance reminiscent of how reconnection-induced diffusivity regulates flux accumulation in phase two of our simulations (Section~\ref{sec:phase2}). For \sgr,~$10^3t_g$ corresponds to a flare period of~${\sim}6 \, \rm h$, of order the characteristic near-IR recurrence timescale \citep{meyer_etal_2009, witzel_etal_2018}. However, this identification is still speculative at this stage; more theoretical work is needed to determine the dependence, if any, of the recurrence time on other aspects such as the dimensionality or the mass, angular momentum, and flux supply from larger scales. 
    \\
    \item Luminosity:

    Given typical horizon-scale magnetic field strengths inferred for \sgr{}, generally in the range~$10{-}100 \, \rm G$ \citep{loeb_waxman_2007, eht_etal_2024, vonfellenberg_etal_2025}, we can use the flux saturation condition,~$\phi = \Phi_{\rm H} / \sqrt{\dot{M}_{\rm H}} \simeq 60$, to compute a characteristic flare luminosity in the context of our model. Since this condition applies in phase two during which mass accretes steadily, we estimate the mass accretion rate using its cycle-averaged value,~$\dot{M}_{\rm H} \simeq \langle \dot{M} \rangle_{\rm cycle}$. Then, to tie this rate to the on-flare electromagnetic heating of particles,~$\langle \dot{\mathcal{E}} \rangle_{\rm flare}$, we use the appropriate efficiency,~$\eta_{\rm flare} = \langle \dot{\mathcal{E}} \rangle_{\rm flare} / \langle \dot{M} \rangle_{\rm cycle} \sim 0.1$, measured in Section~\ref{sec:globalphases} but rounded here to the nearest order of magnitude. Finally, we assume that particles radiate efficiently, so that the observed bolometric luminosity of the flare,~$L$, is close to~$\langle \dot{\mathcal{E}} \rangle_{\rm flare}$. Putting all this together, the flux saturation condition becomes~$\phi = \Phi_{\rm H} / \sqrt{L / \eta_{\rm flare}}$. Rearranging, and using~$\Phi_{\rm H} \simeq 2 \pi B_{\rm H} r_{\rm H}^2$, where~$r_{\rm H} = 2r_g$ for a non-spinning black hole, we obtain
    \begin{align}
        L \sim 10^{35} \, \mathrm{erg}\, \mathrm{s}^{-1} \left( \frac{\eta_{\rm flare}}{0.1} \right) \left( \frac{M_{\rm BH}}{4\times 10^6 \, M_\odot} \right)^{2} \left( \frac{B_{\rm H}}{30 \, \rm G} \right)^{2} \left(\frac{\phi}{60} \right)^{-2} \, .
    \end{align}
    This luminosity is in the observable range of \sgr{} X-ray flares, which occur once per day with a typical luminosity of~$L \sim 10^{34} \, \rm erg\, s^{-1}$ but in extreme cases can reach almost~$L \sim 10^{36} \, \rm erg \, s^{-1}$ \citep{genzel_etal_2010, ponti_etal_2015}. The above estimate would decrease for an imperfect radiative efficiency and probably also for an eruption with finite azimuthal extent -- possible only in 3D.
    \\
    \item Photon energies:
    
    As detailed in Sect.~\ref{sec:ions-accel}, we predict the electron energy distribution to extend up to Lorentz factors of~$\gamma \sim \sigma_{\rm H} m_i/m_e \sim 10 \ m_i/m_e$ before cutting off. Using the proton-electron mass ratio, we expect electron Lorentz factors up to at least~$10^4$. This translates to a synchrotron photon energy of
    \begin{equation}
        \varepsilon_c = \frac{3}{2} \gamma^2 \frac{e\hbar B_{\rm H}}{m_ec} \sim 50 \ \mathrm{eV} \left( \frac{B_{\rm H}}{30 \ \mathrm{G}} \right) \left( \frac{\gamma}{10^4} \right)^2 \, .
    \end{equation}
    Below~${\sim}50\, \textrm{eV}$, reconnection results in a hard electron power-law index, explaining the rising~$\nu F_\nu$ spectra typical of IR \sgr{} flares \citep{ponti_etal_2017, gravity_etal_2021}. Above~${\sim} 50 \, \mathrm{eV}$, the electron distribution may or may not continue. Our 2D simulations potentially impose an artificial cutoff here;~3D studies of reconnection in slab geometry suggest that accelerated particles can extend to much higher energies, owing to the uniquely 3D possibility that particles escape plasmoids to undergo additional acceleration \citep{Zhang2021, Chernoglazov2023}.
    Unless~$\sigma_{\rm H}$ reaches values much higher than in our simulations, such acceleration is necessary in order for particles to reach X-ray emitting energies.
    We should also note that estimates of $B_{\rm H}$ based on one-zone models may be lower than the actual near-horizon value. Indeed, since our simulations show a steep radial dependence of $B^r \propto r^{-2}$, the true local value of $B_{\rm H}$ may exceed observation-based inferences by up to an order of magnitude.
    \\
    \item Astrometry and polarization:

    Our axisymmetric, two-dimensional simulations enforce eruptions to occur simultaneously at all azimuthal angles,~$\phi$, which is in tension with \sgr{} observations. In reality, the \sgr{} flares show orbital motion -- both in real space and in linear polarization -- of a localized hotspot around the black hole \citep{GRAVITY2018}, suggesting an azimuthally restricted emitting region. Our model could be refined to capture these features through two main improvements. First, extending our setup to full 3D would allow for non-axisymmetric eruptions. Second, in order to reproduce orbital motion, our models would need to include angular momentum either embedded in the spacetime (in the form of black hole spin) or into the inflowing plasma.
\end{itemize}

In summary, our simulations reproduce the characteristic duration~(${\sim}30 \, \rm min$) and luminosity of \sgr{} flares. The electron energies and recurrence times that emerge from our models reproduce near-IR flares, but more theoretical work is needed to more robustly constrain the recurrence time. Future 3D simulations may be able to show how particles are accelerated all the way up to X-ray-emitting energies while enabling study of non-axisymmetric eruptions and the formation of orbiting hotspots.

\subsection{Stellar-mass black holes accreting the Galactic ISM}

The Milky Way is expected to host up to~${\sim}10^9$ stellar-mass black holes \citep{shapiro_teukolsky_1983, Agol2002, olejak_2020}, one of which has recently been detected by gravitational microlensing \citep{Sahu2022, lam_etal_2022, lam_lu_2023}. These objects should accrete their surrounding interstellar matter and magnetic field. We present here a discussion of the characteristic luminosities, variability timescales, and peak photon energies expected from such sources were they to accrete via the accretion-eruption mechanism studied here.

As shown in our simulations, the typical duration of an eruption-powered flare is ${\sim}100 t_g = 5 \, \mathrm{ms} \, (M_{\rm BH} /10 M_\odot)$. The recurrence of such flares is not well constrained by our model, as explained in Sect.~\ref{sec:sgrA}, but for the present purposes we speculate that it matches our simulations (and the~\sgr{} near-IR flares, rescaled to a smaller black hole mass). This gives an estimate for the accretion-eruption quasi-period of ${\sim}1000 t_g = 50 \, \mathrm{ms} \, (M_{\rm BH} / 10 M_\odot)$.

Following the discussion of Sect.~\ref{sec:sgrA}, the efficiency with which the black hole converts infalling rest-mass energy into plasma energization, averaged over the full accretion-eruption cycle, is~$\eta_{\rm cycle} = \langle \dot{\mathcal{E}} \rangle_{\rm cycle} / \langle \dot{M} \rangle_{\rm cycle} \sim 0.03$. If the plasma radiates efficiently, the cycle-averaged luminosity\footnote{In Section~\ref{sec:sgrA},~$L$ denotes the luminosity during a flare; here, it denotes the (lower) luminosity averaged over the accretion-eruption cycle.} is~$L \sim \langle \dot{\mathcal{E}} \rangle_{\rm cycle} = 0.03 \langle M \rangle_{\rm cycle}$. Typical accretion rate estimates for isolated stellar-mass black holes feeding on the ISM lie in the range~$10^{10{-}15} \, \rm g \, s^{-1}$ \citep{Agol2002, matsumoto_etal_2018, martinez_etal_2025}, translating to~$L\sim10^{29{-}33} \, \rm erg\, s^{-1}$.

Next, we estimate the magnetic field strength at the horizon of a stellar-mass black hole accreting the ISM. Using the flux saturation condition from phase two of our model,~$\phi = \Phi_{\rm H} / \sqrt{\dot{M}_{\rm H}} \simeq 60$, we estimate that, at the beginning of an eruption,
\begin{equation}
    B_{\rm H} \sim 3\times10^4 \ \mathrm{G} \left( \frac{M_{\rm BH}}{10 M_\odot} \right)^{-1} \left( \frac{\dot{M}_{\rm H}}{10^{10} \textrm{g}\, \textrm{s}^{-1}} \right)^{1/2} \, .
\end{equation}
Such magnetic fields are much stronger than in the case of \sgr{}, amplifying the expected photon energies. Electrons accelerated up to~$\gamma \sim 10^4$, as predicted by our model, radiate synchrotron photons of energy
\begin{align}
    \varepsilon_c = \frac{3}{2} \gamma^2 \frac{e\hbar B_{\rm H}}{m_ec} \sim 20 \ \mathrm{keV} \left( \frac{B_{\rm H}}{10^4 \ \mathrm{G}} \right) \left( \frac{\gamma}{10^4} \right)^2 \, .
    \label{eq:epscsmallbh}
\end{align}
We stress, however (as discussed in Section~\ref{sec:sgrA}), that higher electron energies might be possible in 3D. With a spectrum rising at least until~$\varepsilon_c$, Equation~(\ref{eq:epscsmallbh}) suggests that the spectral peak of isolated black holes lies not in the soft X-rays, but in the hard X-rays and, for the most vigorously accreting sources, beyond. Given current instrumental sensitivities, flares peaking in this range might not be easy to detect \citep[see the recent attempt by][]{mereghetti_etal_2025}. However, given the number of potential sources, their overall population might power a Galactic diffuse hard X-ray background. 

%--------------------------------------------------------------------
\section{Conclusions}\label{sec:Conclusion}

In this work, we have performed self-consistent kinetic simulations that capture a cyclic eruptive activity driven by the accretion of zero-angular-momentum, magnetized plasma onto a Schwarzschild black hole. We show that magnetic reconnection plays a central role in regulating magnetic-flux accretion, triggering magnetospheric eruptions, and accelerating particles to ultrarelativistic energies. Building upon analytical models and empirical knowledge of plasma instabilities, we derive self-consistent constraints on the characteristic particle Lorentz factors and eruption timescales, finding encouraging agreement with observational expectations for \sgr. Because our model is agnostic to the black-hole mass, similar processes may also occur in accreting, isolated stellar-mass black holes, potentially contributing to diffuse galactic high-energy emission.

However, our current setup lacks key features required to reproduce several observational properties of \sgr’s flares, particularly the orbital motion of the emitting region. We expect that incorporating angular momentum -- either as black-hole spin or in the inflowing plasma -- together with fully three-dimensional simulations, will enable us to capture these effects. Such a more realistic framework will also allow for self-consistent modeling of the associated synchrotron emission, including the production of light curves, polarization maps, and spectra, which we leave for future work.

\begin{acknowledgements}
The authors thank I. El Mellah, B. Crinquand, N. Scepi, and K. Kin for insightful discussions. We are thankful for the anonymous referee's comments, which helped improve some of the discussions in the paper.
This project has received funding from the European Research Council (ERC) under the European Union’s Horizon 2020 research and innovation program (Grant Agreement No. 863412).
JM is supported by a grant from the Simons Foundation (MP-SCMPS-00001470).
AS is supported by NSF through grant AST-2508744.
Computing resources were provided by TGCC under the allocations A0170407669 made by GENCI.
\end{acknowledgements}


 \bibliographystyle{aa}
 \bibliography{ref}

\end{document}